\documentclass[fleqn,10pt]{wlscirep}
\usepackage[utf8]{inputenc}
\usepackage[T1]{fontenc}
\usepackage{float}
\usepackage{subfig}
\usepackage{graphicx}
\usepackage{hyperref}
\usepackage{csvsimple}
\usepackage{multirow}
\title{Investor behavior and multiscale cross-correlations: Unveiling regime shifts in global financial markets}

\author[1,*]{Marina Dolfin}
\author[1]{George Kapetanios}
\author[1]{Leone Leonida}
\author[1]{Jose De Leon Miranda}
\affil[1]{King's College London, King's Business School, London, WC2B 4BG, UK}

\affil[*]{marina.dolfin@kcl.ac.uk}


\begin{abstract}
We propose an algorithm to capture emergent patterns in the cross-correlations of financial markets, highlighting regime changes on a global scale. In our approach, financial markets are viewed as complex adaptive systems, and multiscale properties and cross-correlations are considered, particularly during stress conditions such as the COVID-19 pandemic, the invasion of Ukraine by Russia in 2022, and Brexit. We investigate whether significant disruptions reflect an imbalance in investment horizons among investors, and we propose a measure based on this imbalance to depict the impact on global financial markets. The detrended cross-correlation cost (DCCC), which is derived from detrended cross-correlation analysis, uses cross-correlations at different timescales to capture variations in investment horizons amid financial uncertainties. Our algorithm, which combines DCCC analysis and the minimum-spanning-tree filtering approach, tracks system interconnectedness and investor imbalances. We tested the DCCC indicator using daily price series of G7, Russian, and Chinese markets over the past decade and found that it increases sharply during ``crash'' periods compared to ``business as usual'' periods. Our empirical results confirm that short-term investment horizons dominate during financial instabilities; this validates our hypothesis and indicates that the DCCC can serve as a leading indicator of shifts in financial-market regimes.
\end{abstract}

\makeatletter\def\Hy@Warning#1{}\makeatother
\usepackage{microtype}

\begin{document}

\flushbottom
\maketitle

\thispagestyle{empty}

\section*{Introduction}
In attempting to understand the complex structures of financial markets, it is helpful to view them as ecosystems \cite{May:2008nature,Haldane:2011nature}, which are traditionally modeled as complex adaptive systems. Like ecosystems, financial markets exhibit complexity and adaptability, and just as ecosystems can reach critical thresholds leading to significant shifts, so too can financial markets undergo sudden systemic failures known as regime shifts. Critical transitions \cite{Scheffer:2009nature} appear to be related to emergent patterns in cross-correlations among the fundamental entities of the system. This observation leads to the possibility of exploring the network of spatial cross-correlation dynamics in global financial markets in relation to regime transitions, resilience, and, possibly, controllability. If the global market network can be considered as a complex dynamical system, one may consider the investors as the main agents acting at the microscopic level and their heterogeneous investment horizons as principal characteristics.

Our research seeks to identify patterns indicative of financial instabilities that may trigger regime shifts. The tool developed in this work analyzes time-varying nonlinear cross-correlations among market indices spanning various timescales. By examining how these correlations change over time, the project seeks to better understand the dynamics of the market and how they relate to the activities of both short- and long-term investors. This approach recognizes that market volatilities at different time resolutions reflect the influences of traders operating on varying timescales, thus providing insights into overall market behavior \cite{Muller:1997jef}. The main modelling approach is based on tracking observations using a rolling-window technique; this includes examining the interrelationships among different levels of observation---from pairwise correlations to system-wide connectedness---in relation to both short- and long-term timescales. We hypothesize that using a multiscale approach across observation levels can provide critical insights into the stability of the complex dynamical system under examination.

In an unstable market, investors with very short horizons are assumed to be the most active, and some long-term investors may adopt shorter investment horizons \cite{kristoufek2013}. At critical points, long-term investors will restrict or even stop their trading activities, and short-term investors will become dominant. The existence and activity of investors with a wide range of investment horizons are essential; they contribute to the smooth and stable functioning of the market. Herein, we investigate whether, during periods of market instability or in the preceding periods, short-term cross-correlation coefficients deviate significantly from long-term cross-correlation coefficients in comparison to those during stable periods \cite{Scheffer:2009nature}.

Our conjecture is based on the following reasoning \cite{kristoufek2012}. In a regular market, the scaling of variance should be stable, and fluctuations for different horizons should lie on a straight line. If any of the investment horizons becomes dominant, the scaling of variance will be less precise, meaning fluctuations for different horizons might deviate from a stable pattern. As stated in Ref.~\citeonline{kristoufek2013}, when the investment horizons are evenly distributed, the market functions efficiently (as supply and demand are met). The proposal to use the detrended cross-correlation coefficients at different scales to determine the weights of a minimum spanning tree (MST) for real-time regime-switch identification is innovative. Monitoring the evolution of the MST, updated at consistent intervals, appears crucial for identifying significant shifts in network topology that could signal regime changes. Incorporating different timescales enhances the depth of our analysis; this not only enables the detection of behavioral shifts among investor classes over time, particularly between short- and long-term investors, but also captures the dual nature of market dynamics, encompassing both transient shocks and more lasting systemic changes. A significant rise in long-term cross-correlations might indicate a greater alignment among long-term investors in their responses to market events, suggesting a potential shift in market sentiment or underlying trends.

To assess the spread of trading activity across different investment horizons, we compare the normalized tree length of the MST, which is constructed by linking stocks based on the detrended cross-correlation analysis (DCCA) coefficient, which is transformed into a Euclidean distance \cite{mantegna1999,mantegna2003}. Specifically, we define a ratio---the detrended cross-correlation cost (DCCC)---between the normalized length of the MST at a shorter scale and that at a longer one. The greater the fluctuations of this ratio, the less stable the scaling, and thus the less stable the market.

In summary, we propose the DCCC as an indicator of financial instability, taking into account the implications of heterogeneous investor time horizons and the typical feedback loops of social and financial systems \cite{Scheffer:2009nature}. We also explore patterns that may arise when changes in macroeconomic variables impact investors' time horizons. We tested our conjecture using daily share prices of market indices as proxies for the G7, Chinese, and Russian markets, demonstrating the applicability of our approach to real-world financial data.

\section*{Results}
\subsection*{Monitoring the time evolution of the financial-market network}
We use daily time series of closing prices for proxies of the G7, Chinese, and Russian markets, including the S\&P~500, S\&P/TSX Composite Index, CAC~40, DAX Performance Index, FTSE MIB Index, Nikkei~225, FTSE~100, Hang Seng Index, and the IMOEX Russia Index. The time series span approximately 10~years, from 5 March 2013 to 30 May 2023. We quantify correlations using the DCCA coefficient, filtered through a GARCH(1,1) model estimate, as introduced by Kristoufek \cite{kristoufek2014} (see \hyperref[Methods]{Methods} section).

\begin{table*}[ht]
\centering
\begin{tabular}{|c|c|c|c|c|c|c|c|c|c|}
\hline
\multicolumn{1}{|c|}{\multirow{2}{*}{\textbf{}}} & \multicolumn{9}{c|}{\textbf{DCCA coefficients ($s=1$~month)}} \\ \cline{2-10}
\multicolumn{1}{|c|}{} & \textbf{US} & \textbf{CA} & \textbf{FR} & \textbf{DE} & \textbf{IT} & \textbf{JP} & \textbf{UK} & \textbf{CN} & \textbf{RU} \\ \hline
\textbf{US} & 1 & & & & & & & & \\ \hline
\textbf{CA} & 0.8280 & 1 & & & & & & & \\ \hline
\textbf{FR} & 0.7336 & 0.7491 & 1 & & & & & & \\ \hline
\textbf{DE} & 0.7350 & 0.7269 & 0.9416 & 1 & & & & & \\ \hline
\textbf{IT} & 0.6543 & 0.6786 & 0.8995 & 0.8697 & 1 & & & & \\ \hline
\textbf{JP} & 0.6050 & 0.5814 & 0.6624 & 0.6429 & 0.5874 & 1 & & & \\ \hline
\textbf{UK} & 0.6960 & 0.7548 & 0.8667 & 0.8215 & 0.7727 & 0.6054 & 1 & & \\ \hline
\textbf{CN} & 0.4948 & 0.4886 & 0.5283 & 0.5085 & 0.4645 & 0.5446 & 0.5302 & 1 & \\ \hline
\textbf{RU} & 0.3948 & 0.4495 & 0.4649 & 0.4677 &0.4328 & 0.3520 &0.4584 & 0.3497 & 1 \\ \hline
\end{tabular}
\caption{DCCA coefficients computed across the entire dataset using a 1-month timescale.}
\label{tab:1month}
\end{table*}

\begin{table*}[ht]
\centering
\begin{tabular}{|c|c|c|c|c|c|c|c|c|c|}
\hline
\multicolumn{1}{|c|}{\multirow{2}{*}{\textbf{}}} & \multicolumn{9}{c|}{\textbf{DCCA coefficients ($s=4$~months)}} \\ \cline{2-10}
\multicolumn{1}{|c|}{} & \textbf{US} & \textbf{CA} & \textbf{FR} & \textbf{DE} & \textbf{IT} & \textbf{JP} & \textbf{UK} & \textbf{CN} & \textbf{RU} \\ \hline
\textbf{US} & 1 & & & & & & & & \\ \hline
\textbf{CA} & 0.8584 & 1 & & & & & & & \\ \hline
\textbf{FR} & 0.7972 & 0.7986 & 1 & & & & & & \\ \hline
\textbf{DE} & 0.7913 & 0.7745 & 0.9321 & 1 & & & & & \\ \hline
\textbf{IT} & 0.6950 & 0.7287 & 0.9128 & 0.8625 & 1 & & & & \\ \hline
\textbf{JP} & 0.7212 & 0.6405 & 0.7217 & 0.7567 & 0.6484 & 1 & & & \\ \hline
\textbf{UK} & 0.7869 & 0.8280 & 0.8734 & 0.8429 & 0.7786 & 0.6742 & 1 & & \\ \hline
\textbf{CN} & 0.4972 & 0.4731 & 0.5104 & 0.4895 & 0.4717 & 0.4481 & 0.5716 & 1 & \\ \hline
\textbf{RU} & 0.4788 & 0.4997 & 0.5387 & 0.5514 &0.5290 & 0.4510 &0.5257 & 0.3614 & 1 \\ \hline
\end{tabular}
\caption{DCCA coefficients computed across the entire dataset using a 4-month timescale.}
\label{tab:4months}
\end{table*}

Tables~\ref{tab:1month} and \ref{tab:4months} list static estimates of the DCCA coefficients across the whole date range for 1- and 4-month timescales, respectively. We then non-linearly transform the DCCA coefficients to convert the correlations into Euclidean distances\cite{mantegna1999, mantegna2003}; this uses the advantages of an appropriate taxonomy and of a positive adjacency matrix (see \hyperref[Methods]{Methods} section). Hereinafter, we refer to this measure as the DCCA distance.

\begin{figure}[H]
\centering
\includegraphics[width=\textwidth]{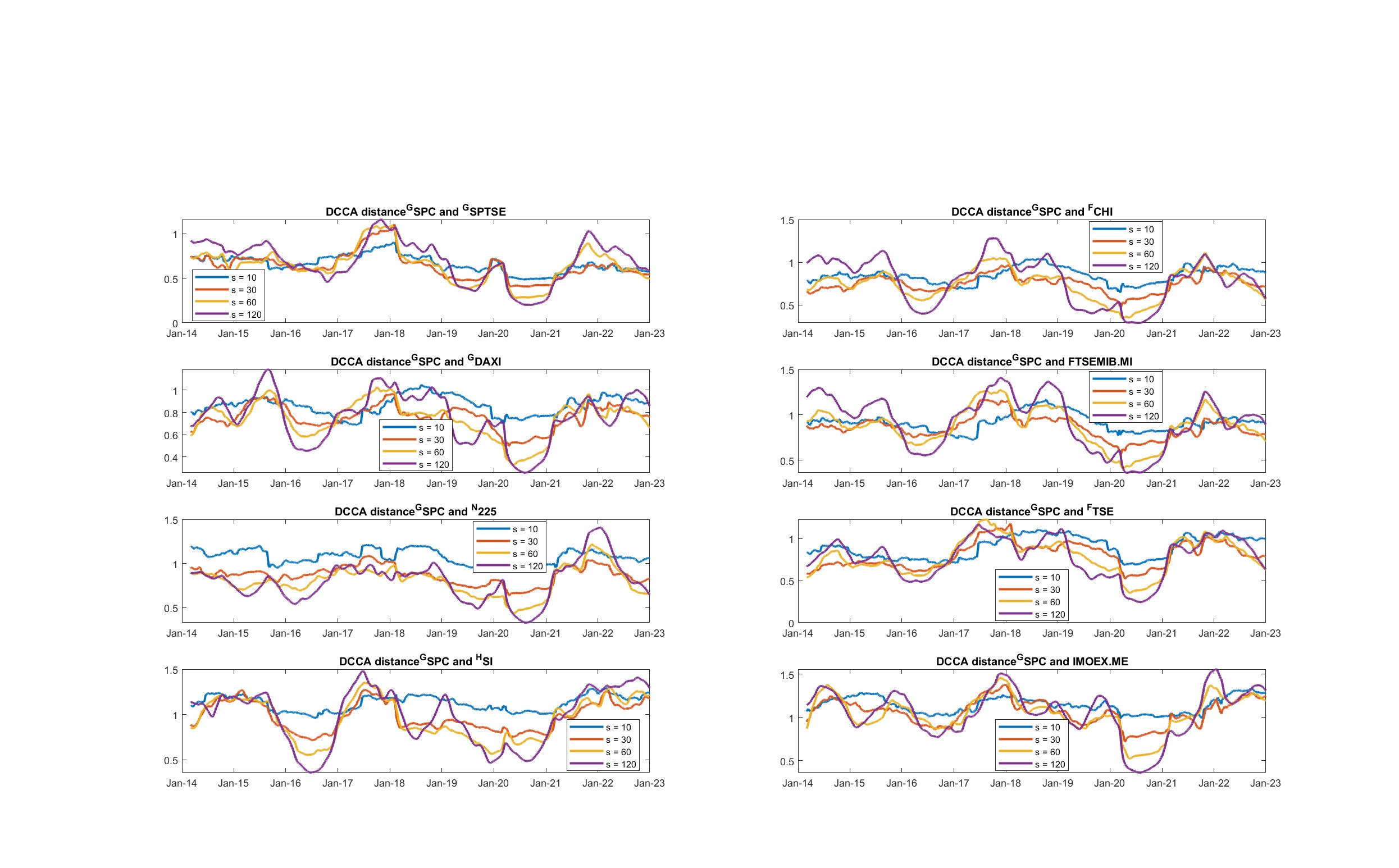} 
\caption{DCCA distances between S\&P 500 index and other indices, computed for different timescales ranging from s = 10 to 120 days, with a rolling window of w = 250 days.}
\label{fig:paircorrelations}
\end{figure}

When analyzing the time evolution of the DCCA distances for different pairs of market indexes, the DCCA correlation distance between, for example, the S\&P index and other indices generally tends to be lower during periods of crisis and higher during periods of stability, as illustrated in Fig.~\ref{fig:paircorrelations}. Data suggest that the correlation distances between the S\&P index and other indices tend to be smaller at longer timescales (e.g., 60~days) and higher at shorter timescales (e.g., 5~days). At the same time, for longer time horizons, for example after the COVID-19 pandemic, uncertainties were dissipated, and the DCCA distances for longer time horizons increased more significantly and more rapidly, mainly with Western indices. This may suggest that the rebalancing of portfolios of long-term investors may be pronounced both when a crisis starts and when it recedes. Furthermore, the smaller distances on longer scales during a crisis could indicate that long-term investors, such as institutions or pension funds, are impacted more significantly. This could be due to their long-term investment strategies, which typically involve larger and more established companies that are potentially more exposed to the global impacts of an event such as the COVID-19 pandemic. The significant increase in long-scale distances as COVID-19-related doubts dissipate may suggest that these investors see the emerging recovery as an opportunity to diversify their portfolios, reducing cross-correlations. Conversely, the fact that the short-term DCCA distances remain relatively low even as the crisis eases may suggest that short-term investors, such as day or swing traders, continue to trade in a more correlated manner. This could be because their trading strategies are more influenced by short-term market sentiment and technical indicators, which can still respond to volatile market conditions.

\begin{figure}[ht!]
\centering
{{\includegraphics[width=7cm]{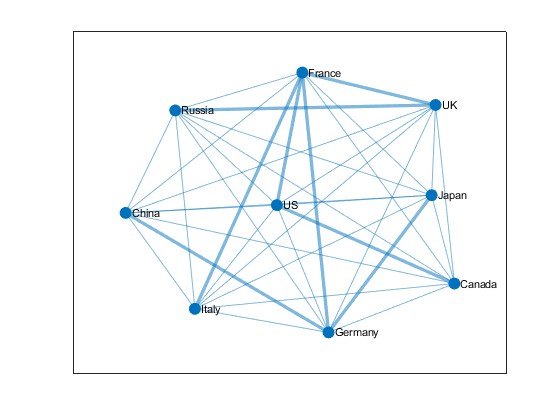} }}
\qquad
{{\includegraphics[width=7cm]{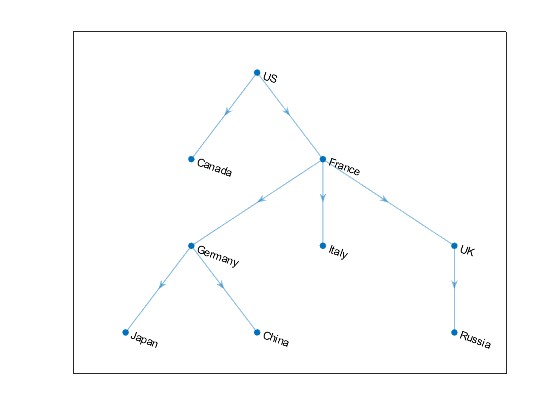} }}
\caption{Pre-COVID-19 network (left), showing data from 5 March 2013 to 5 December 2019, and its corresponding MST (right), with $w=250$~days and $s=120$~days.}
\label{fig:preCovid}
\end{figure}

\begin{figure}[ht!]
\centering
{{\includegraphics[width=7cm]{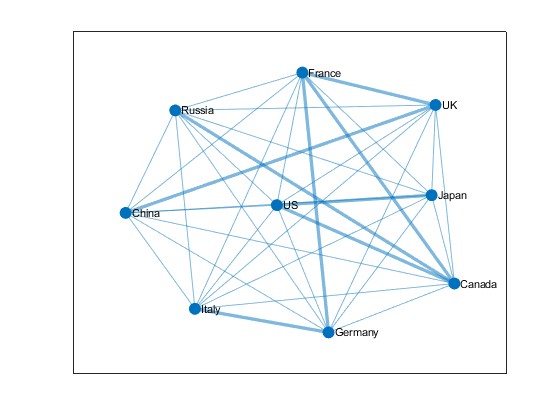} }}
\qquad
{{\includegraphics[width=7cm]{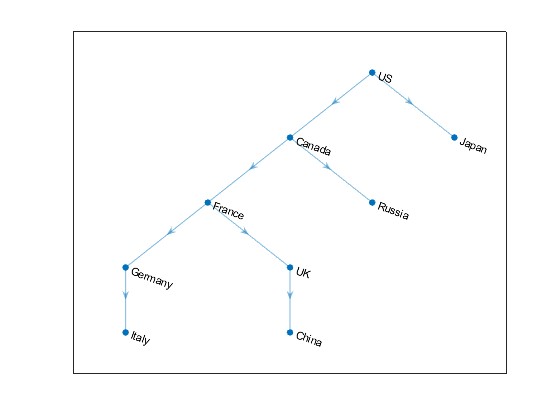} }}
\caption{Post-COVID-19 network (left), showing data from 5 December 2021 to 30 May 2023, and its corresponding MST (right), with $w=250$~days and $s=120$~days.}
\label{fig:postCovid}
\end{figure}

In our approach, we construct a complete undirected weighted network, in which each node represents a financial market and each weighted link is determined by the DCCA distance between the two nodes. The dynamics of the network are tracked using a rolling-window approach. Specifically, an edge between nodes $i$ and $j$ is measured as $d^{ij}_{\mathrm{DCCA}}(s,t)=\sqrt{2(1-(\rho^{ij}_{\mathrm{DCCA}}(s,t)^2)}$ using the DCCA coefficient $\rho^{ij}_{\mathrm{DCCA}}(s,t)$ ($i,j=1,\dots,N$), where $N$ is the number of nodes, $t$ represents time, and $s$ is the timescale (see \hyperref[Methods]{Methods} section). The resulting DCCA distances are then collected into an $N\times N$ time-dependent adjacency matrix for the evolving network. We apply a filtering approach using the MST. The MST is a subset of the network comprising a connected, edge-weighted, undirected graph that connects all the vertices without any cycles while minimizing the sum of edge weights; this plays a crucial role in our analysis, as it provides information about the structure of the network, particularly during market instabilities. Figures~\ref{fig:preCovid} and \ref{fig:postCovid} show two graphical examples of specific network configurations and their corresponding MSTs, before and after the onset of COVID-19, respectively.

\begin{figure}[ht!]
\centering
{{\includegraphics[width=9cm]{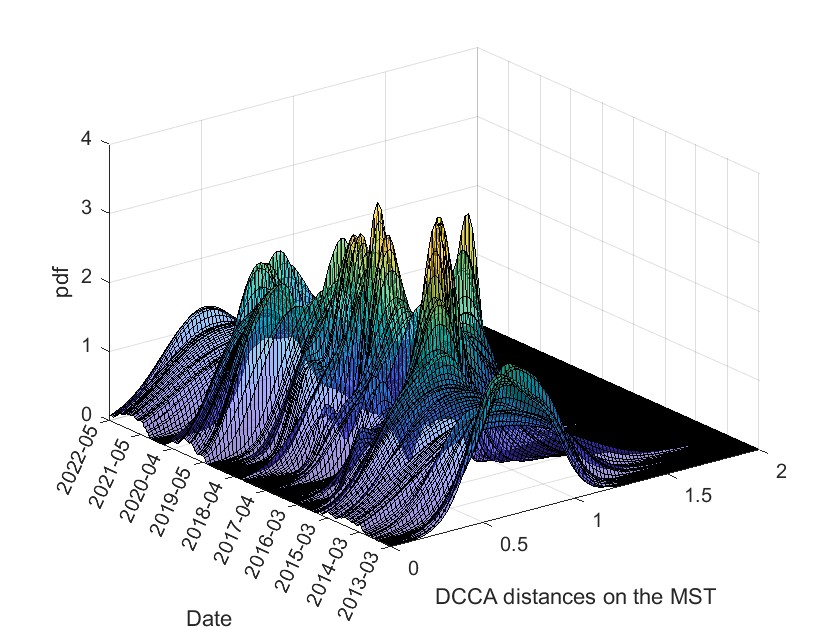} }}
\qquad
{{\includegraphics[width=7cm]{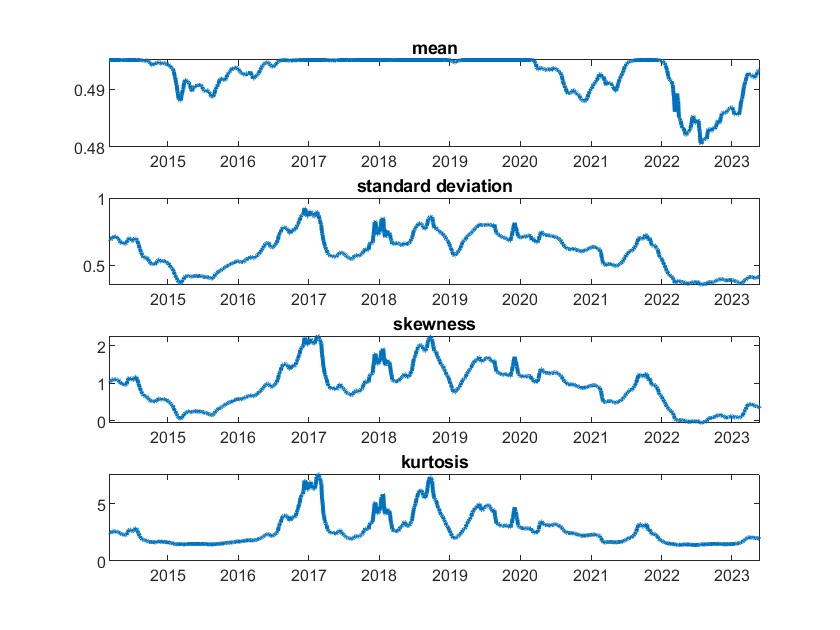} }}
\caption{Time evolution of the density function of the 1-month DCCA distances filtered on the MST (left) and the corresponding dynamics of the first four moments (right).}
\label{fig:MST30}

\end{figure}

\begin{figure}[ht!]
\centering
{{\includegraphics[width=9cm]{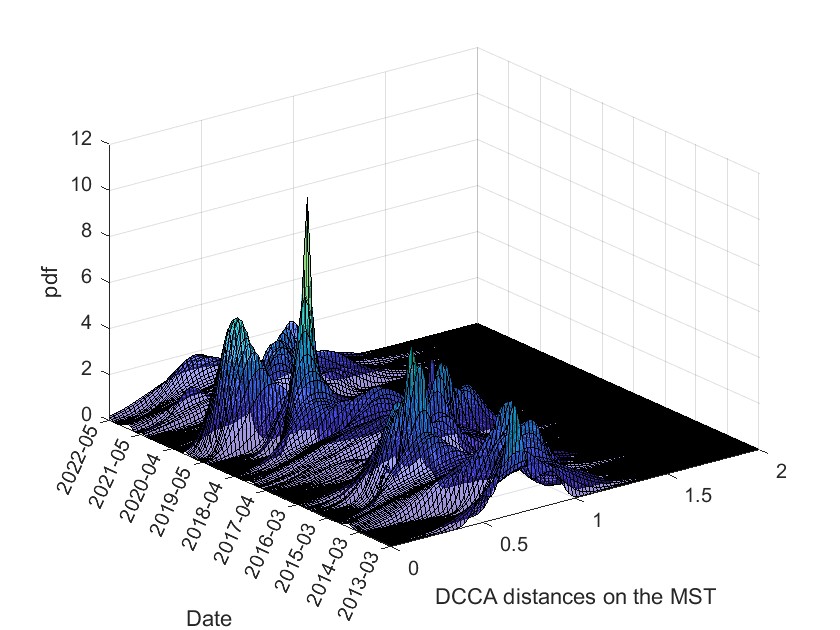} }}
\qquad
{{\includegraphics[width=7cm]{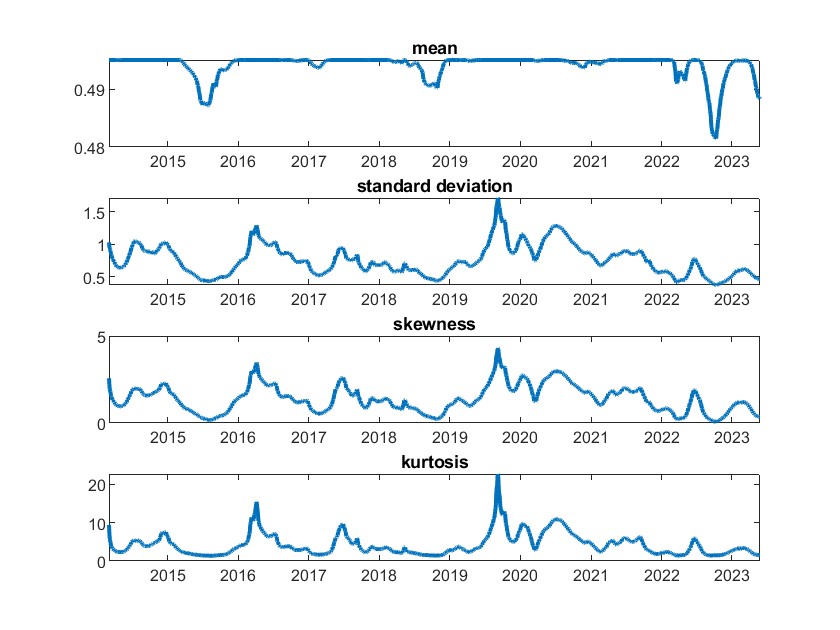} }}
\caption{Time evolution of the density function of the 4-month DCCA distances filtered through the MST (left) and the corresponding dynamics of the first four moments (right).}
\label{fig:MST120}
\end{figure}

During market instabilities, the tree has a shorter than average length and is generally tightly packed. The MST highlights the primary structure (skeleton) of a complex system with non-trivial dynamics and serves as a reduced representation of the entire cross-correlation matrix, containing essential information about correlations. Figures~\ref{fig:MST30} and \ref{fig:MST120} show the estimated densities of DCCA distances filtered through the MST for 1- and 4-month timescales, respectively, using a kernel distribution object by fitting it to the data computed with a rolling window of 1~year of trading days ($w = 250$~days). The corresponding first four moments are plotted on the right of each figure.

\begin{figure}[ht!]
\centering
{{\includegraphics[height= 7cm,width=7cm]{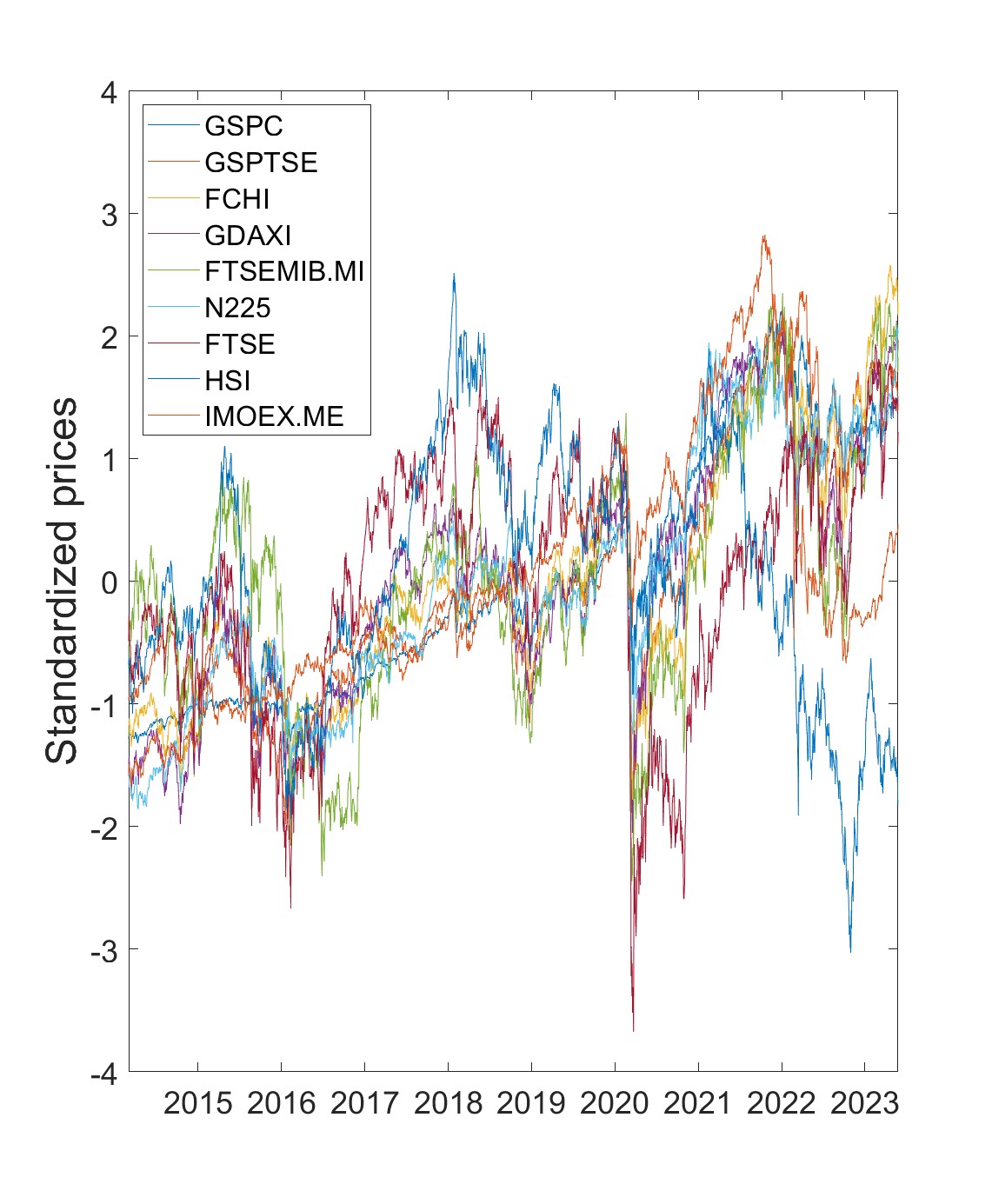} }}
\qquad
{{\includegraphics[height=7cm,width=7cm]{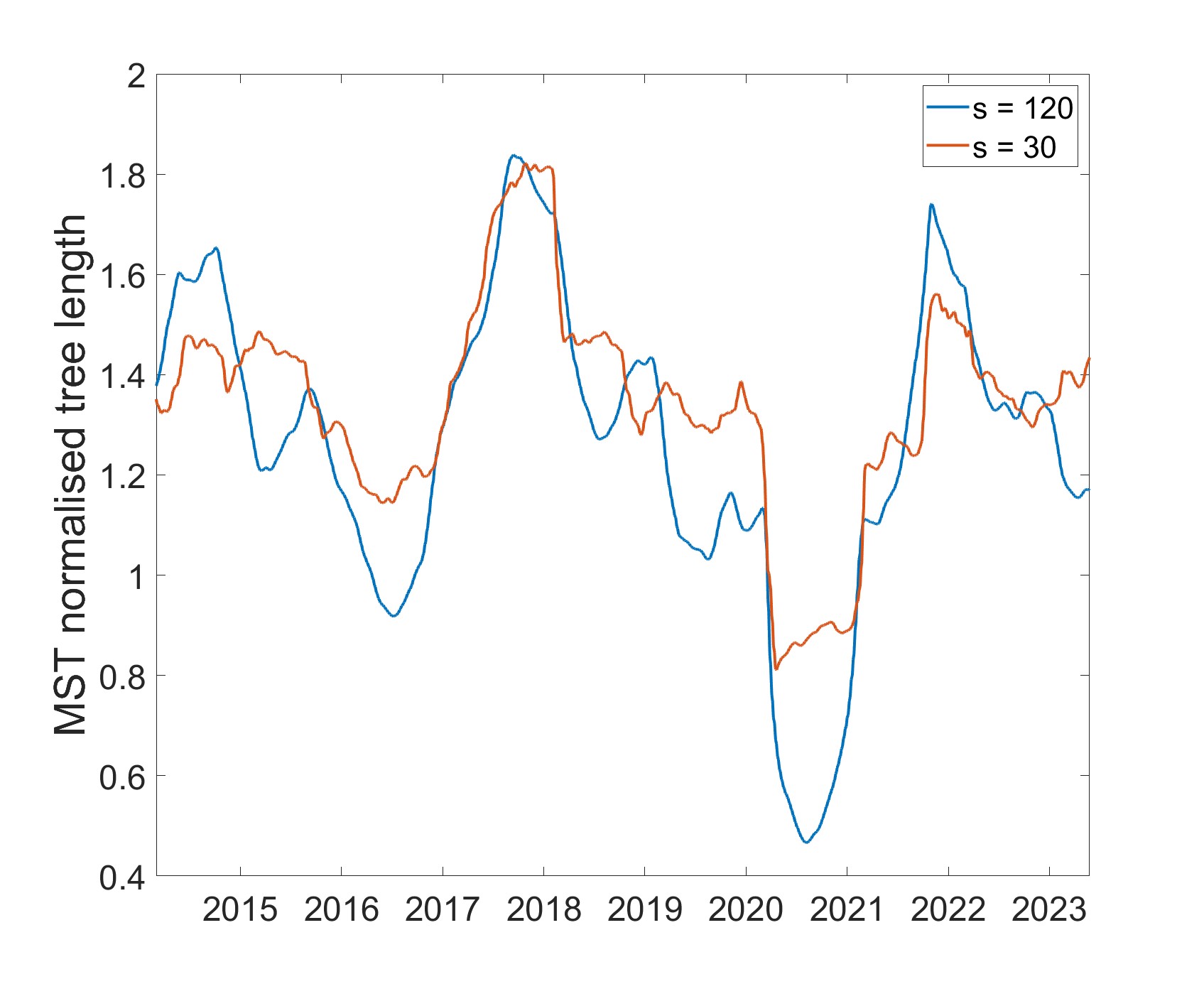} }}
\caption{Standardized prices for the nine proxies of financial markets (left) and the MST normalized tree length for $s=1$~month and $s=4$~months (right) with a rolling window of $w=250$~days.}
\label{fig:prices}
\end{figure}

Monitoring the evolution of the MST structure as it is updated at consistent intervals for different timescales could allow identification of notable shifts in network topology; this might be crucial, as such shifts could indicate regime changes. Here, we specifically seek to monitor the imbalances between the MST structures for different timescales (1~month versus 4~months). Figure~\ref{fig:prices} shows the standardized prices of the nine proxies of the G7, Chinese, and Russian markets, and the average tree lengths of the DCCA distances filtered through the MST, computed for timescales of 1 and 4~months.

\subsection*{DCCC and spectral graph analysis}
\begin{figure}[ht!]
\centering
{{\includegraphics[width=7cm]{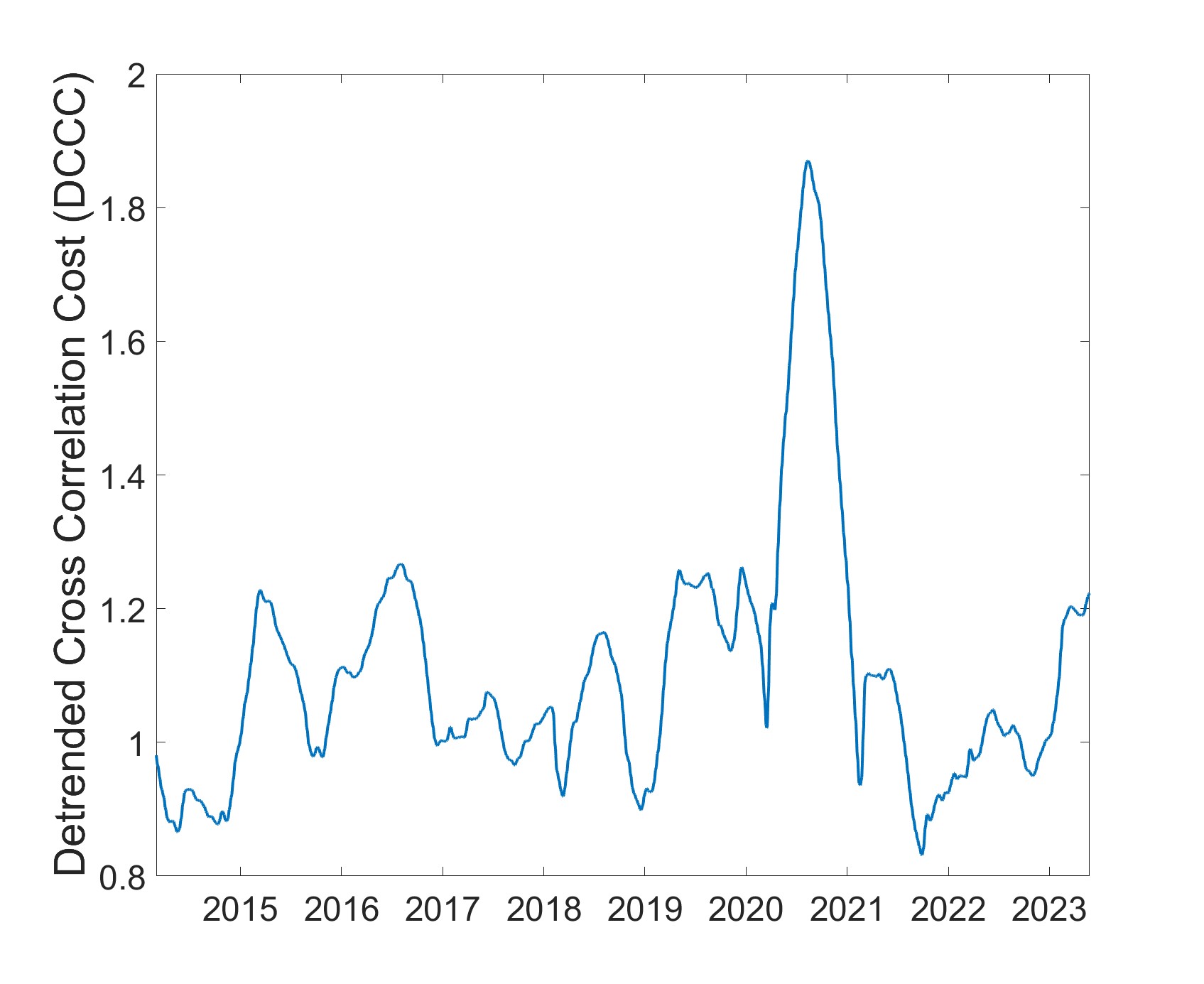} }}
\qquad
{{\includegraphics[width=7cm]{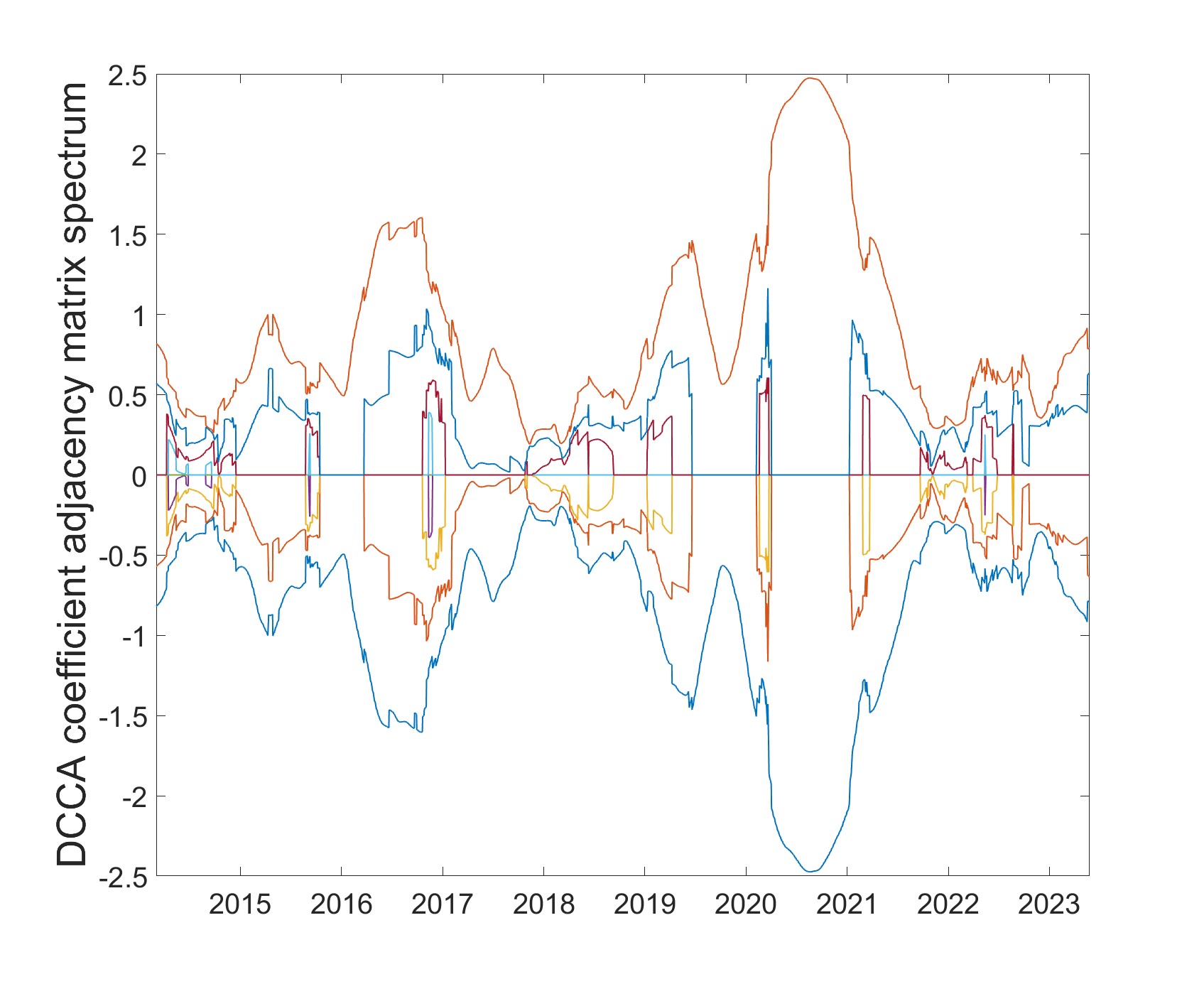} }}
\caption{Dynamics of the 4-month/1-month DCCC (left) and the adjacency-matrix spectrum with a 4-month timescale (right) with $w=250$~days.}
\label{fig:ratio}
\end{figure}

We now relate the time evolution of the DCCC to the evolution of the dominant eigenvalue of the time-evolving adjacency matrix of the MST constructed using the DCCA coefficients. Figure~\ref{fig:ratio} shows the evolution of the DCCC (with a rolling window $w = 250$~days) and of the spectrum of the adjacency matrix of the network defined by the detrended cross-correlations, using a long-term scale ($s = 120$~days). In our framework, the adjacency matrix of cross-correlation distances is positive, so Perron's theorem applies. The dominant eigenvalue of the adjacency matrix $\lambda_{\max}$ is thus simple and has a positive eigenfunction; $\lambda_{\max}$ can be bounded through crucial network quantities \cite{sarkar2011spectral}, specifically $\langle k\rangle \leq \lambda_{\max} \leq k_{\max}$, where $k$ is the network degree. Perron's theorem ensures that an iterative process will converge when the dominant eigenvalue is at most 1 \cite{Perron}. Moreover, the financial interpretation of the dominant eigenvalue is the so-called ``market mode'' (see Ref.~\citeonline{potters2005financial} and references therein). The dominant eigenvalue, normalized by the average volatility of the stocks, is considered to be a proxy for the average correlation between the stocks. The dominant eigenvalue of the adjacency matrix and the ratio of the two average tree lengths ($s=30$ and $120$~days)---i.e., the DCCC---exhibit strong correlation, with a Pearson correlation coefficient ranging from 0.82 to 0.90 when using a rolling window that spans $w = 150$ to $500$~days. Through spectral graph theory (see Ref.~\citeonline{Perron} and references therein), it can be shown that the dominant eigenvalue is related to a higher level of interconnectedness, implying systemic risk and financial instability \cite{Podobnik:2012}.

Our conclusion is that the ratio of the two MST average tree lengths (the DCCC) shows promise as an indicator for identifying patterns indicative of financial instability due to its consistently high correlation with the dominant eigenvalue of the adjacency matrix of cross-correlations. We intend to use this indicator---appropriately rescaled between 0 and 1---as the transition probability in a Markov process within a regime-switching model. Further details will be discussed in the \hyperref[Research perspectives]{Research perspectives} section.

\subsection*{Sensitivity analysis}
\begin{figure}[ht!]
\centering
{{\includegraphics[width=7cm]{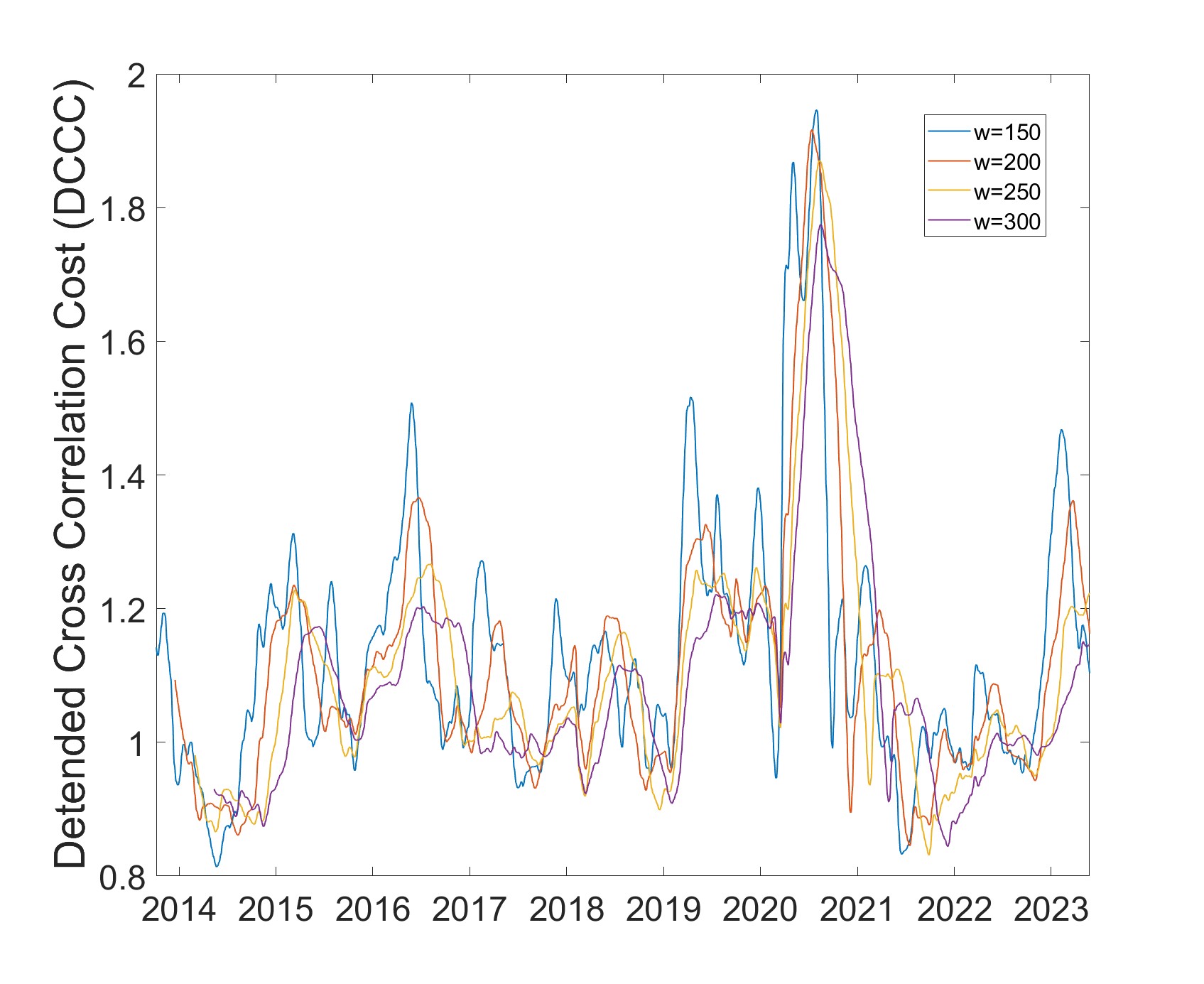} }}
\qquad
{{\includegraphics[width=7cm]{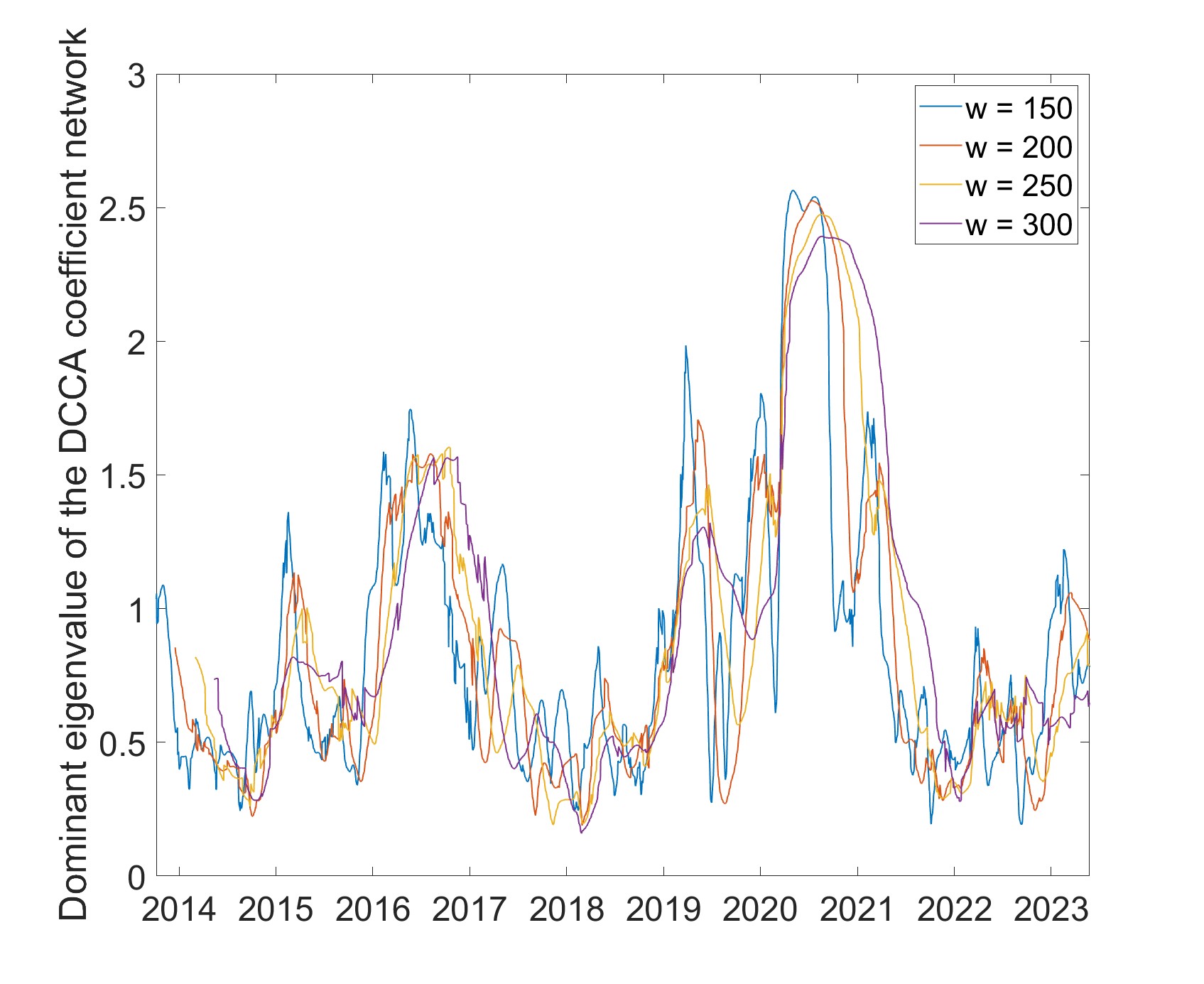}}}
\caption{DCCC values (left) and the dominant eigenvalue of the adjacency matrix of the MST network of DCCA coefficients (right) for different rolling-window sizes.}
\label{fig:ratios}
\end{figure}

We conducted various robustness checks. Figure~\ref{fig:ratios} shows the DCCC and the long-term eigenvalues ($s=120$~days) plotted for different rolling-window sizes ranging from 150 to 300~days. It can be seen that the DCCC and the dominant eigenvalue consistently exhibit high correlations, with Pearson correlation coefficients ranging from 0.82 to 0.84 for rolling windows between $w=150$ and $300$~days; these correlations become even stronger for larger rolling windows, reaching up to 0.90 for $w=500$~days. These findings indicate that pruning DCCA distances that change over time with the MST is highly effective for generating meaningful distances that have potential utility in various applications, including but not limited to portfolio management \cite{Onnela2003}.

\subsection*{Comparison with the DCCA-GARCH approach of Diebold et~al.}
Our methodology can be compared with the variance-decomposition technique discussed by Diebold et~al. \cite{Diebold2014} and Diebold and Yilmaz \cite{Diebold2023}; as a brief comparison with our proposed measures, in this section, we develop the Diebold et~al. connectedness table, which displays pairwise directional connectedness from stocks $j$ to $i$ and also total directional connectedness as the off-diagonal row and column sums. In the work of Diebold et~al., connectedness refers to the extent to which shocks or unexpected changes can spread from one firm, sector, or country to another in a financial system. The measure is designed to quantify the potential for systemic risk, or the risk of a broad financial-system breakdown.

To measure connectedness, Diebold et~al. used variance decompositions from vector autoregressions; these allow the variance of an error term in one variable to be attributed to shocks in that variable and other variables. In the context of Ref.~\citeonline{Diebold2014}, the variance of a firm's stock return can be decomposed into parts that are due to shocks to that firm itself and shocks to other firms. They then construct a network using these variance decompositions, where the nodes are firms and the edges are the contributions of one firm's shocks to the variance of another firm's returns. This involves breaking down the forecast error variance of variable $i$ into components attributed to various variables in the system. Specifically, $d^H_{ij}$ represents the $ij$th $H$-step variance-decomposition component. This quantifies the fraction of the variance in the forecast error of the $i$th variable over $H$ steps that can be attributed to shocks in variable $j$. Each element $d^H_ij$ in the $N\times N$ block of the connectedness table represents the contribution of variable $j$'s shock to the $H$-step-ahead forecast error variance of variable $i$. All measures of connectedness---whether they examine pairwise or system-wide relationships---rely on cross-variance decompositions denoted as $d^H_{ij}$ ($i,j = 1, \dots, N$; $i \neq j$), where the off-diagonal entries represent the ``non-own'' or ``cross'' variance decompositions. The focus is then on the quantities for which $i\neq j$. In addition, a grand total of off-diagonal entries is provided. The model assumes correlated shocks.

The authors of Ref.~\citeonline{Diebold2014} use the generalized variance decompositions (GVDs), as introduced by Koop et~al.\cite{koop1996impulse} and further developed by Pesaran and Shin\cite{pesaran1998generalized}; this provides an alternative method for computing variance decompositions in vector autoregression (VAR) models when the shocks are correlated. Unlike the Cholesky-based approach, GVDs do not require a specific ordering of the variables; they treat each variable as if it is ``first in the ordering'' while allowing shocks to be correlated and accounting for correlations among shocks based on observed historical data. Unlike Cholesky factorization, which attempts to orthogonalize the shocks, GVDs maintain the correlation structure of the shocks.

\begin{figure}[H]
\centering
\includegraphics[height=8cm,width=10cm]{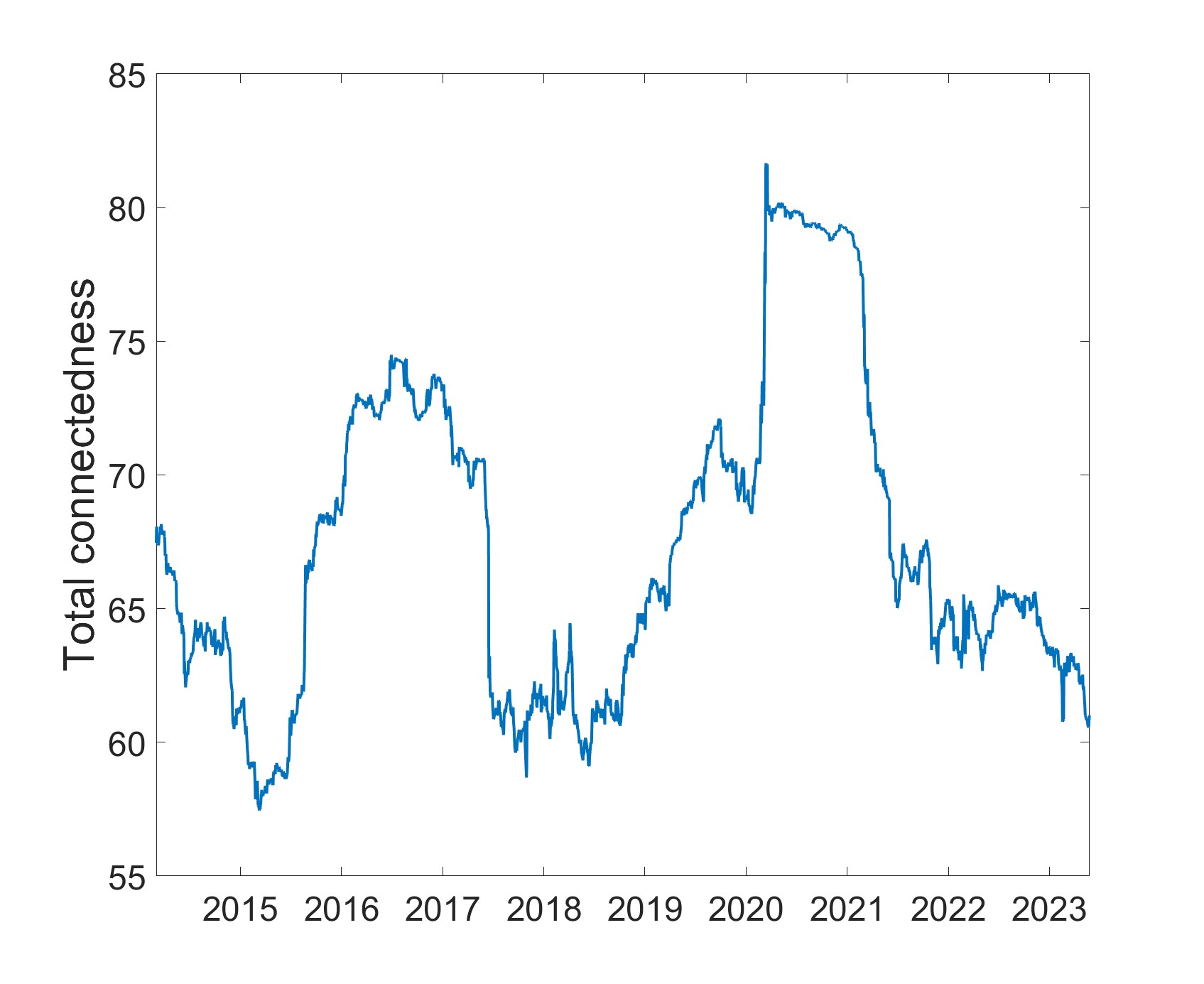}
\caption{Time series of Diebold et~al. total connectedness with $w=250$~days.}
\label{fig:Dieboldconn}
\end{figure}

In Refs.~\citeonline{Diebold2014} and \citeonline{Diebold2023}, a rolling-window approach was used to track the time-varying connectedness in real time. This means that for each point in time, a VAR model is estimated using only the most recent $w$ observations. This process is then repeated for each point in time. Following this approach, the time series of total directional connectedness (``to'', ``from'', and ``net'' degrees) can be evaluated. Figure~\ref{fig:Dieboldconn} shows a time series of total directional connectedness with $w=250$~days as obtained using the Diebold et~al. measure of connectedness with the same data as in the previous sections. A general total directional table is provided in \hyperref[app:b]{Appendix~B}.

This measure of total connectedness is highly correlated with the dominant eigenvalues of the adjacency matrix of the network with links constructed with the DCCA coefficients (specifically a Pearson correlation coefficient of 0.81 with rolling window $w=250$~days and similar high positive coefficients with different rolling-window widths). We have chosen to incorporate this different measure into the algorithm for the purpose of comparison, and the computations for the ``to'', ``from'', and ``net'' degrees are included in the code provided in the repository linked at the end of this article.

\section*{Discussion}
We emphasize the nuanced behavior of investors with different time horizons and highlight the multifaceted benefits of monitoring cross-correlations for various stakeholders in financial markets. When comparing long- and short-term horizons, one needs to consider that long-term scales reflect enduring market drivers, encompassing macroeconomic shifts and extended industry trends, while short-term investors respond to daily market variations and are influenced by news, macroeconomic events, or immediate trading approaches. Transitioning from the dynamics of investment horizons, the slow-herding phenomenon that is characteristic of long-term investors implies that they do not generally respond impulsively to daily news, but major events can trigger collective portfolio reassessments, leading to more synchronized investment decisions over longer durations \cite{Sias}.

Turning our attention to investor behavior in response to disruptions, significant market upheavals can increase uncertainty about future economic trajectories, prompting both short- and long-term investors to pivot to more secure assets. Long-term investors might display pronounced shifts as they seek to rebalance portfolios in anticipation of prolonged economic unpredictability; regulatory factors, such as compliance with specific investment criteria, can influence this behavior \cite{caballero2008}. It is thus noted that continuous monitoring of cross-correlations between stock markets can facilitate effective portfolio diversification and risk management for various stakeholders.

The contributions of our research are as follows. First, the DCCA coefficient is transformed into a proper metric, and DCCA distances can be used in the context network theory to efficiently explore transitions between market regimes \cite{Schweitzer2009}. A novel perspective with respect to Refs.~\citeonline{Diebold2014} and \citeonline{Diebold2023} is also presented regarding network tools: the transformation of the cross-correlations estimations into a practical metric allows us to explore the evolving MST structure; moreover, in estimating cross-correlations, we relax the assumption of covariance stationarity implied by the Diebold et~al. approach. We also present a novel perspective on the work of Billio et~al. (Refs.~\citeonline{Billio2012} and \citeonline{Billio2023} and references therein) with respect to Granger causality; specifically, the standard definition of Granger causality is linear, and they introduce a nonlinear Granger causality based on a Markov-switching model of asset returns. However, their assumptions are quite restrictive: all the relevant information from the past history of the process at the current time is represented only by the previous state; moreover, they use a Markov chain with stationary transition probabilities. The more recent paper by Billio et~al.\cite{Billio2023} extends the classical capital asset pricing model by including network linkages, obtaining a network-augmented linear factor model. This network is directed but not weighted, so the size of the cross-correlations are not explicitly taken into account.

The logic of our approach---linking the estimate of correlations with network theory by transforming correlations into measures---was introduced in the seminal work of Mantegna et~al. \cite{mantegna1999, mantegna2003} and a large amount of subsequent literature based on it. Nonetheless, to the best of our knowledge, this approach has not previously been applied to the case of nonlinear cross-correlations.

\subsection*{Research perspectives}
\label{Research perspectives}
We propose using the DCCA distances and the DCCC indicator at varying scales to determine the weights in the MSTs of market networks for real-time regime-switch identification. Incorporating different timescales not only allows for the detection of behavioral shifts among investor classes over time---notably between short- and long-term investors---but it also captures the dual nature of market dynamics, transient shocks, and more lasting systemic changes. Furthermore, the observation of sudden shifts in the balance between short- and long-term investors, as observed using the DCCC, can act as a leading indicator of different regimes.

We present a list summary of research perspectives as follows:
\begin{itemize}
\item Regarding the modeling of regime changes, in our work, the parameters of an autoregression are viewed as the outcome of a discrete-state Markov process. The important paper by Hamilton \cite{hamilton1989} addresses optimal probabilistic inferences regarding regime shifts based on the observed behavior of the gross national product, and the turning point is a structural event that is inherent in the data-generation process. We thus propose using the DCCC to define the transition probabilities of the Markov process.

\item Emergent phenomena of synchronization in networks should be explored as the emergence of a coherent state in the global system. The question of whether the onset of synchronization is related to critical points should be examined.

\item The adjacency matrix of DCCA distances could be used for portfolio optimization.

\item The proposed approach could be used to explore the prevalence of interconnectedness across groups with respect to interconnectedness within groups, where groups are market indices (such that ``within groups'' means analyzing connectedness across single stocks enclosed in the market index). This approach allows the use of multi-layered network-modelling techniques.

\end{itemize}

\section*{Methods}
\label{Methods}
\subsection*{Data collection and transformation}
We used daily recorded time series of the closing prices of proxies for the markets of the G7 countries (namely GSPC, GSPTSE, FCHI, DAX Performance Index, FTSEMIB, N225, FTSE), along with those for China (HSI) and Russia (IMOEX). Data were collected from 5 March 2013 to 30 May 2023.

For the general $i$th index ($i=1,\dots,N$ and $N=9$) we computed the logarithmic returns of the standardized prices as:
\begin{equation}
r_i(t)= \ln\, \tilde P_i(t)-\ln\,\tilde P_i(t-1),\quad t=1,\dots,T,
\end{equation}
where prices are measured with a given frequency (daily in our example) and are standardized into time series with zero mean and unit variance. Specifically,
\begin{equation}
\tilde P_i(t)=\frac{P_i(t)-\langle P_i(t)\rangle }{\sigma_P},
\end{equation}
in which $\sigma_P=\sqrt{\langle P^2\rangle -\langle P\rangle ^2}$ is the standard deviation of the time series of prices $P_i(t)$ ($i=1,\dots,T$), and $\langle \dots \rangle $ denotes a time average over the study period.

\subsection*{Measuring time-dependent cross-correlation distances across different timescales}
The detrended cross-correlation distances are obtained by transforming the DCCA coefficient $\rho_{\mathrm{DCCA}}(s)$ into a Euclidean distance, where $s$ represents the chosen timescale \cite{Podobnik:2008}. The DCCA coefficient is defined as the ratio between the detrended covariance $F^2_{\mathrm{DCCA}}(s)$ of two time series \cite{Peng1994} and the product of the detrended variance functions (DFAs) $F_{\mathrm{DFA},x}(s)$ and $F_{\mathrm{DFA},y}(s)$ for the pair of time series being analyzed, expressed by:
\begin{equation}
\rho_{\mathrm{DCCA}}(s)=\frac{F^2_{\mathrm{DCCA}}(s)}{F_{\mathrm{DFA},x}(s)F_{\mathrm{DFA},y}(s)}.
\end{equation}
The detrended variance for a time series is obtained as follows. First, the time series ${x}$ is integrated, yielding $X_t=\sum_{i=1}^t(x_i-\langle x\rangle)$ ($t=1,\dots,T$). Then, one constructs $n$ mutually exclusive boxes of equal length $T/n=s$ and quantifies the local trend $\hat X_n(t)$ for each as the ordinate of a linear least-squares fit.

The detrended time series is obtained by subtracting the local trend from the original time series. The DFA function $F_{\mathrm{DFA},x}(s)$ is then computed by taking the square root of the average, over all boxes, of the detrended time series. This is a function of the length of each box, i.e., the timescale.

The detrended covariance $F^2_{\mathrm{DCCA}}(s)$ \cite{Podobnik:2008} is obtained using a similar procedure. First, the time series ${x_i}$ and ${y_i}$ are integrated; then, each is detrended within each box using a linear least-squares fit, giving $X_n(t)=X(t)-\hat X_t$ and $Y_n(t)=Y(t)-\hat Y_t$. For each box, the covariance is
\begin{equation}
f^2_{\mathrm{DCCA}}=\frac 1{n-1}\sum_{t=i}^{i+n}X_n(t)Y_n(t).
\end{equation}
The detrended covariance $F^2_{\mathrm{DCCA}}(s)$ is ultimately computed by averaging on all the boxes.

We compute both the detrended covariance and the detrended variance by filtering the series, specifically by estimating a conditional variance with a GARCH(1,1) model and applying the transformation $r_{t,f}=r_t/\sqrt{h_t}$, where $h_t$ represents a conditional variance obtained from a GARCH(1,1) model. The GARCH filter reduces the potential volatility bias between different time windows. Similar to the linear Pearson coefficient, the DCCA coefficient is bounded such that $-1\leq \rho_{\mathrm{DCCA}}(s)\leq 1$ for each timescale $s$.

Following reasoning analogous to that in Ref.~\citeonline{mantegna1999}, the DCCC distance between two time series is obtained by transforming the DCCA coefficient into a Euclidean distance:
\begin{equation}\label{transformation}
d^{ij}_{\mathrm{DCCA}}(s) = \sqrt{2(1 - (\rho^{ij}_{\mathrm{DCCA}}(s))^2) }
\end{equation}
where $i$ and $j$ refer to two nodes in the network. The nonlinear transformation (\ref{transformation}) is applied for each instant of time in our time range.

The scale dependence helps with understanding the trade-off between long- and short-term features. Kristoufek~\cite{kristoufek2014} showed that the DCCA coefficient dominates the Pearson coefficient for non-stationary series. The critical values for the DCCA coefficient can be found in Ref.~\citeonline{Podobnik:2011}.

To track time-varying connectedness, we use a uniform estimation window of width $w$, sliding through the sample with one-day steps; i.e., we compute $d^{ij}_{\mathrm{DCCA}}(s,t)$ using observations in the interval $[t,w+t-1]$ with $t=1,\dots,T-w+1$, where $T$ is the sample size.

\subsection*{Detrended cross-correlation cost (DCCC)}
We construct an undirected time-dependent weighted network with $N$ market indices as nodes and the time-dependent cross-correlation distances as weighted links. Using Prim's algorithm, we construct the MST, denoted by $\mathbf T^t$. The MST is a simple connected graph that connects all $N$ nodes of the graph with $N-1$ links such that the sum of all the weights of the links is a minimum. If the weights are positive, as in our framework, the MST is a minimum-cost subgraph \cite{spingGlass}.

We use the normalized tree length as a measure of the degree of connectivity:
\begin{equation}
L(s,t)=\frac 1{N-1}\sum_{d^{ij}_{\mathrm{DCCA}}\in T^t}d^{ij}_{\mathrm{DCCA}}(s,t).
\end{equation}
The normalized tree length provides a measure of the network's efficiency. The DCCC tracks the ratio of the normalized tree length on a shorter timescale to that on a longer one over time:
\begin{equation}
DCCC(s_1,s_2,t)=L(s_1,t)/L(s_2,t),
\end{equation}
where $s_1 < s_2$.

\section*{Data and code availability}
All data, along with the complete code, which includes automatic access to the Yahoo Finance database, are available from \href{https://github.com/JoseDLM/Investor-Behavior-and-Multiscale-Cross-Correlations}{https://github.com/JoseDLM/Investor-Behavior-and-Multiscale-Cross-Correlations}. This can be easily updated to run with different stocks over various time ranges or with different data frequencies.

\section*{Appendix A: Statistics of interest}
\label{app:a}
\setcounter{table}{0}
\renewcommand{\thetable}{A\arabic{table}}
Table~\ref{tab:statistics} presents some statistics of interest, including annualized mean return, annualized volatility, skewness, kurtosis, autocorrelation at different time intervals, and autocorrelation of squared returns.

\begin{table}[htbp]
\centering
\resizebox{\textwidth}{!}{
\begin{tabular}{lccccccccc}
\toprule
Stat\textbackslash Tickers& GSPC & GSPTSE & FCHI & GDAXI & FTSEMIB.MI & N225 & FTSE & HSI & IMOEX.ME \\
\midrule
Ann mean & 0.10 & 0.04 & 0.06 & 0.07 & 0.05 & 0.09 & 0.02 & $-$0.02 & 0.06 \\
Ann vol & 0.18 & 0.15 & 0.19 & 0.19 & 0.23 & 0.20 & 0.16 & 0.20 & 0.22 \\
Skewness & $-$0.82 & $-$1.66 & $-$0.81 & $-$0.58 & $-$1.41 & $-$0.19 & $-$0.88 & 0.04 & $-$2.25 \\
Kurtosis & 19.24 & 45.30 & 13.49 & 12.85 & 19.01 & 7.77 & 15.94 & 6.90 & 53.33 \\
AC 1d & $-$0.14*** & $-$0.09*** & $-$0.00 & $-$0.01 & $-$0.06*** & $-$0.02 & $-$0.00 & 0.01 & 0.09*** \\
AC 5d & 0.05*** & 0.04*** & 0.00 & 0.01 & 0.01*** & $-$0.01*** & 0.02 & $-$0.03 & $-$0.02*** \\
AC 10d & $-$0.06*** & $-$0.03*** & $-$0.03*** & $-$0.04*** & $-$0.01*** & $-$0.02*** & $-$0.00*** & $-$0.00 & $-$0.04*** \\
AC 20d & $-$0.02*** & 0.00*** & 0.02** & 0.03*** & $-$0.00** & $-$0.01*** & 0.04*** & 0.05* & $-$0.02*** \\
AC sq ret 1d & 0.48*** & 0.43*** & 0.11*** & 0.06*** & 0.15*** & 0.21*** & 0.17*** & 0.28*** & 0.68*** \\
AC sq ret 5d & 0.32*** & 0.28*** & 0.12*** & 0.08*** & 0.06*** & 0.08*** & 0.11*** & 0.13*** & 0.02*** \\
AC sq ret 10d & 0.24*** & 0.14*** & 0.11*** & 0.09*** & 0.05*** & 0.14*** & 0.14*** & 0.09*** & 0.01*** \\
AC sq ret 20d & 0.11*** & 0.06*** & 0.05*** & 0.07*** & 0.01*** & 0.03*** & 0.09*** & 0.06*** & 0.02*** \\
\bottomrule
\multicolumn{10}{l}{\small Notes: *, **, and *** indicate significance at the 0.05, 0.01, and 0.001 levels, respectively.}\\
\end{tabular}
}
\caption{\label{tab:statistics}Summary statistics.}
\end{table}

\section*{Appendix B: Diebold et~al. measure of connectedness}
\label{app:b}
\setcounter{table}{0}
\renewcommand{\thetable}{B\arabic{table}}

Table~\ref{tab:example} lists the values of the Diebold et~al. measure of connectedness calculated in this work.

\begin{table}[htbp]
\centering
\resizebox{\textwidth}{!}{
\begin{tabular}{|l|*{10}{c|}}
\hline
\multicolumn{1}{|c|}{} & \textbf{GSPC} & \textbf{GSPTSE} & \textbf{FCHI} & \textbf{GDAXI} & \textbf{FTSE.MIB} & \textbf{N225} & \textbf{FTSE} & \textbf{HSI} & \textbf{IMOEX.ME} & \textbf{From} \\ \hline
\textbf{GSPC} & 27.59 & 18.18 & 10.89 & 10.56 &8.83 & 6.93 & 9.93 &4.28 & 2.76 & 72.40 \\ \hline
\textbf{GSPTSE} & 18.34 & 27.39 & 18.86 & 10.02 & 9.41 & 4.71 & 11.75 & 3.87 & 3.60 & 72.60 \\ \hline
\textbf{FCHI} & 8.30 & 8.74 & 20.97 & 18.33 & 16.40 & 4.67 & 15.29 & 3.68 & 3.59 & 79.02 \\ \hline
\textbf{GDAXI} & 8.46 & 8.42 & 18.92 & 21.83 & 15.99 & 4.82 & 14.45 & 3.48 & 3.58 & 78.16 \\ \hline
\textbf{FTSE.MIB} & 7.82 & 8.53 & 17.98 & 16.96 & 23.98 & 4.42 & 13.53 & 3.19 & 3.53 & 76.01 \\ \hline
\textbf{N225} & 5.69 & 5.52 & 7.42 & 6.43 & 5.52 & 51.29 & 7.39 & 8.55 &2.15 & 48.70 \\ \hline
\textbf{FTSE} & 8.37 & 10.28 & 16.76 & 15.35 & 13.58 & 4.18 & 23.28 & 4.11 & 4.04 & 76.71 \\ \hline
\textbf{HSI} & 3.99 & 5.25 & 7.60 & 6.80 & 5.62 & 6.37 & 8.43 & 52.54 & 3.35 & 47.45 \\ \hline
\textbf{IMOEX.ME} & 4.63 & 6.52 & 8.40& 8.08 & 7.47 & 2.71 & 8.56 & 3.28 & 50.31 & 49.68 \\ \hline
\textbf{To} & 65.64 & 71.47 & 98.88 & 92.56 & 82.86 & 38.85 & 89.37 & 34,47 & 26.64 & 66.75 \\ \hline
\textbf{Net} & $-$6.76 & $-$1.13 & 19.85 & 14.39 & 6.85 & $-$9.84 & 12.65 & $-$12.97 & $-$23.03 & 0 \\ \hline
\end{tabular}
}
\caption{\label{tab:example}Diebold et~al. connectedness values obtained in this work.}
\end{table}

\bibliography{sample}

\section*{Author contributions statement}
All authors contributed equally to this work. M.D., G.K., L.L. and J.DLM. were all involved in the design of the study, conducting the experiments, analyzing the data, interpreting the results, and writing the manuscript. All authors reviewed and approved the final manuscript.
\section*{Additional information}
Competing Interests:
The authors declare no competing interests.

Data Availability:
The code used to produce the numerical results and to generate the data in this study is available in the GitHub repository, \href{https://github.com/JoseDLM/Investor-Behavior-and-Multiscale-Cross-Correlations}{https://github.com/JoseDLM/Investor-Behavior-and-Multiscale-Cross-Correlations}.
\end{document}